# Electronic Structure Descriptor for Discovery of Narrow-Band Red-Emitting Phosphors

Zhenbin Wang[1], Iek-Heng Chu[1], Fei Zhou,[2] Shyue Ping Ong[1]★

[1] Department of NanoEngineering, University of California San Diego, 9500 Gilman Drive, La Jolla, CA 92093-0448, United States

[2] Physics Division, Lawrence Livermore National Laboratory, 7000 East Avenue, Livermore, CA 94550, United States

**Narrow-band red-emitting phosphors are a critical component in phosphor-converted light-emitting diodes for highly efficient illumination-grade lighting. In this work, we report the discovery of a quantitative descriptor for narrow-band $Eu^{2+}$-activated emission identified through a comparison of the electronic structure of known narrow-band and broad-band phosphors. We find that a narrow emission bandwidth is characterized by a large splitting of more than 0.1 eV between the two highest $Eu^{2+}$ $4f^7$ bands. By incorporating this descriptor in a high throughput first principles screening of 2,259 nitride compounds, we identify five promising new nitride hosts for $Eu^{2+}$-activated red-emitting phosphors that are predicted to exhibit good chemical stability, thermal quenching resistance and quantum efficiency, as well as narrow-band emission. Our findings provide important insights into the emission characteristics of rare-earth activators in phosphor hosts, and a general strategy to the discovery of phosphors with a desired emission peak and bandwidth.**

Phosphor-converted light-emitting diodes (pc-LEDs) are an energy efficient and environmentally friendly lighting source for solid-state lighting.[1–7] For illumination-grade lighting, a warm white light LED with a high luminous efficacy and colour rendering index is desirable. A common strategy to improve the colour rendering index is to add a red-emitting component into LEDs. In recent years, $Eu^{2+}$-activated nitride phosphors have emerged as highly promising red-emitting phosphors for pc-LEDs due to their high chemical and thermal stability, small thermal quenching and high quantum efficiency (QE).[8–10] Unfortunately, the broad bandwidth and/or deep red maximum of the emission spectra of commercial red-emitting phosphors lead to a substantial portion of the emitted energy being outside the range of human vision, severely reducing the luminous efficacy of the LED. For example, though the widely used commercial $CaAlSiN_3:Eu^{2+}$ phosphor has an excellent QE > 95% at 150 °C,[1] its emission peak ($\lambda_{em}$ ~ 650 nm)[2] is too red-shifted and its bandwidth (characterized by the full width at half maximum, or FWHM ~ 90 nm) is too broad.

Recently, Pust et al.[5] reported the discovery of a highly promising narrow-band red-emitting phosphor $SrLiAl_3N_4:Eu^{2+}$ with an emission peak position at ~ 650 nm and a FWHM of ~ 50 nm. A 14% improvement in luminous efficacy was achieved with $SrLiAl_3N_4:Eu^{2+}$ as a red-emitting component in a pc-LED compared with a commercial LED device.[5] The luminous efficacy will be further enhanced if the emission maximum of $SrLiAl_3N_4:Eu^{2+}$ could be blue-shifted, but the activator concentration is found to have little influence on the emission band position. The chemically similar $CaLiAl_3N_4:Eu^{2+}$ narrow-band red-emitting phosphor also has an emission peak that is too deep red ($\lambda_{em}$ ~ 668 nm) for high-powered LEDs.[6] Another narrow-band red-emitting phosphor SrMg$_3$SiN$_4$:Eu$^{2+}$ has an ideal peak position at ~ 615 nm and a narrow bandwidth of ~ 43 nm.[7] However, the thermal quenching of SrMg$_3$SiN$_4$:Eu$^{2+}$ is too severe for practical applications. Despite these discoveries, the overall number of known narrow-band red-emitting phosphors remains very small. The development of highly efficient narrow-band red-emitting phosphors with an optimal spectral peak position ($\lambda_{em} \sim 615$ nm) and a narrow width (FWHM < 50 nm) is therefore a critical materials challenge for warm white illumination-grade lighting.[11,12]

In this work, we report the discovery of a quantitative descriptor for narrow-band Eu$^{2+}$-activated emission that emerged from a comparison of the first principles electronic structure of nine well-known phosphors. We then identified five new narrow-band red-emitting phosphors, CaLiAl$_3$N$_4$ ($P\bar{1}$), SrLiAl$_3$N$_4$ ($I4_1/a$), SrLiAl$_3$N$_4$ ($P\bar{1}$), SrMg$_3$SiN$_4$ ($P\bar{1}$), and BaLiAl$_3$N$_4$ ($P\bar{1}$), for high-power pc-LED applications through a high-throughput screening of ternary and quaternary nitride compounds. These five new phosphors are predicted to satisfy a balance of chemical stability, good thermal quenching behaviour and quantum efficiency, and narrow-band emission.

The emission spectrum of Eu$^{2+}$-activated phosphors is attributed to the $4f^65d^1 \rightarrow 4f^7$ electronic transition. We therefore begin with a hypothesis that narrow-band and broad-band emission can be identified by differences in the electronic band structure. In this work, we adopt the definition of having a measured FWHM < 60 nm as a narrow-band emitter. The calculated electronic band structures of two representative red-emitting phosphors, SrLiAl$_3$N$_4$:Eu$^{2+}$ (narrow-band) and CaAlSiN$_3$:Eu$^{2+}$ (broad-band), are shown in Figures 1a and b respectively. We find that the Eu$^{2+}$ 4$f$ bands are very narrowly distributed in energy, and the energy level of each band may be estimated by averaging across all $k$-points. Figure 1c shows the average Eu$^{2+}$ 4$f$ band levels of five narrow-band (SrLiAl$_3$N$_4$:Eu$^{2+}$, SrMg$_3$SiN$_4$:Eu$^{2+}$, CaLiAl$_3$N$_4$:Eu$^{2+}$, BaLi$_2$Al$_2$Si$_2$N$_6$:Eu$^{2+}$ and Si$_{141}$Al$_3$ON$_{191}$(β-SiAlON):Eu$^{2+}$)[5–7,13,14] and four broad-band (CaAlSiN$_3$:Eu$^{2+}$, Sr$_2$Si$_5$N$_8$:Eu$^{2+}$, Ba$_2$Si$_5$N$_8$:Eu$^{2+}$ and SrAlSi$_4$N$_7$:Eu$^{2+}$)[2,15,16] phosphors derived from their calculated band structures (Supplementary Figure S1). With the exception of the green-emitting BaLi$_2$Al$_2$Si$_2$N$_6$:Eu$^{2+}$ and β-SiAlON:Eu$^{2+}$, all the remaining phosphors exhibit red emission. Interestingly, we find that all narrow-band phosphors, regardless of their emission colour, have a distinct large energy splitting $\Delta E_S$ of > 0.1 eV between the two highest Eu$^{2+}$ 4$f$ bands. Broad-band emitters, on the other hand, have a more uniform 4$f$ band distribution, with several bands lying within 0.1 eV of the highest 4$f$ band at the Fermi level.

We may provide a justification for this distinct narrow-band electronic structure feature by considering the fact that broad-band emission is the result of multiple overlapping emission spectra occurring within the red spectra region. Multiple emission spectra can be the result of either multiple distinct activator sites with similar doping energies,[1,15] or multiple transitions even within the same activator site. Figure 2 shows a schematic of the different transitions in a broad-band and narrow-band phosphor. In both cases, the main emission is the result of the transition of an electron from the lowest 5$d$ band into the empty top 4$f$ band, i.e., a 4$f^6$5$d^1$ → 4$f^7$ transition. However, when there are other 4$f$ bands within 0.1 eV of the top band, transitions to lower 4$f$ bands may occur, resulting in overlapping emissions within ~30 nm of the main peak and broadening of the emission. Therefore, $\Delta E_S$ > 0.1 eV is a descriptor for narrow-band emission.

For commercial applications, a red-emitting phosphor host must also satisfy a demanding set of other properties besides narrow-band emission, namely good phase stability, emission in the red-orange region of the visible spectrum, excellent thermal quenching resistance and high photoluminescence quantum efficiency. The phase stability of a material can be estimated by calculating the energy above the linear combination of stable phases in the phase diagram, also known as the energy above hull $E_{hull}$. Stable materials have an $E_{hull}$ of 0 eV and the more unstable a material, the higher the $E_{hull}$. The emission wavelength has a direct relationship to the band gap $E_g$ of the host material as the conduction band minimum of the host sets the energy level of the excited $Eu^{2+}$-activated phosphor (see Supplementary Figure S2).[17] In general, host materials that are rigid and have large photoionization energy,[17] defined as the energy gap between the conduction band minimum of the host and the lowest 5$d$ levels of the activator, exhibit higher quantum efficiencies, especially at elevated temperatures. The rigidity of a crystal can be estimated by its Debye temperature $\Theta_D$,[18] and the host band gap $E_g$ may be used as a proxy to estimate the photoionization energy. The details of the first principles calculations for each of these properties are given in the Methods section.

We first parameterized the criteria for our high-throughput screening of nitrides through an analysis of the $E_g$ and $\Theta_D$ of ten well-known red-emitting phosphor hosts, summarized in Table 1. Unsurprisingly, the $E_g$ calculated using the Perdew-Burke-Ernzerhof (PBE)[19] functional underestimates the experimental $E_g$ by around 28 ~ 38% due to the well-known self-interaction error and lack of derivative discontinuity in semi-local exchange-correlation functionals.[20] The screened hybrid Heyd-Scuseria-Ernzerhof[21,22] (HSE) functional yields $E_g$ that are in much better agreement (within 0.3 eV) with the experimental values.

The key observation from Table 1 is that despite the systematic underestimation of $E_g$ by PBE, relative trends in $E_g$ are generally well-reproduced, consistent with previous first principles studies.[23] Hence, an efficient screening of $E_g$ may be carried out using the computationally inexpensive PBE functional for a large number of compounds, followed by a more accurate secondary screening with the expensive HSE functional for a more limited subset of compounds. From Table 1, we find that the PBE and HSE $E_g$ for red-emitting phosphors lie in the range of 2.42-3.58 eV and 3.68-4.76 eV, respectively. We also observe that the known phosphor nitrides have relatively high $\varTheta_D$ of more than 500 K, with the exception of $SrSiN_2$ and $BaSiN_2$. Indeed, $SrSiN_2$ and $BaSiN_2$ have non-rigid layered crystal structures,[24] and the $Eu^{2+}$-activated hosts show relatively poor quantum efficiencies of 25% and 40%.[25] For $E_{hull}$, previous successful high-throughput screening efforts[26] have shown that an upper threshold of 50 meV / atom yields materials that are reasonably synthesizable, and the experimentally-known phosphor hosts in Table 1 have $E_{hull}$ between 0 meV / atom and 31 meV / atom. We will therefore adopt $E_{hull}$ < 50 meV / atom as the stability threshold in this work.

Equipped with the narrow-band descriptor as well as the parameterized screening criteria, we carried out a high-throughput first principles screening of ternary and quaternary nitride compounds for narrow-band red-emitting phosphor hosts. Candidate host materials were obtained from all existing ternary and quaternary nitride compounds in the Materials Project database.[27] To further expand the dataset, we also performed a prediction of new nitridosilicate and nitridoaluminate structures with formula $A_xB_yC_zN_n$ (A = Ca/Sr/Ba, B = Li/Mg, C = Al/Si) by applying a data-mined ionic substitution algorithm[28] on all crystal structures in Inorganic Crystal Structure Database.[29] This focused structure prediction effort is motivated by the fact that nitridosilicate and nitridoaluminate hosts have thus far demonstrated the greatest promise as narrow-band red emitters. In total, 2,259 (203 ternary, 156 quaternary and 1,900 predicted nitridosilicate and nitridoaluminate quarternary structures) materials were evaluated on their phase stability ($E_{hull}$), emission wavelength and thermal quenching resistance ($E_g$ and $\Theta_D$), and emission bandwidth ($\Delta E_S$).

A total of eight narrow-band, red-emitting phosphor hosts were identified from our high-throughput screening. Their calculated properties are summarized in Table 2, and their band structures and crystal structures are provided in the Supplementary Information. The calculated properties of 40 other hosts that satisfy all screening criteria with the exception of $\Delta E_S > 0.1$ eV, *i.e.,* they are predicted to be broad-band red-emitting phosphors, are given in Supplementary Table S1. Where there are multiple distinct sites for $Eu^{2+}$ doping, the relative energies and phosphor properties based on $Eu^{2+}$-activation on each distinct site are reported. Three of the identified materials, $CaLiAl_3N_4$ ($I4_1/a$), $SrLiAl_3N_4$ ($P\bar{1}$), and $SrMg_3SiN_4$ ($I4_1/a$) have already been previously reported experimentally as narrow-band red-emitting phosphors, which provides a good validation of our screening approach.

Five of the identified hosts are new materials that have not been previously reported as narrow-band red-emitting phosphors. All five new phosphor hosts belong to three structural prototypes. $CaLiAl_3N_4$ ($P\bar{1}$), $SrMg_3SiN_4$ ($P\bar{1}$), and $BaLiAl_3N_4$ ($P\bar{1}$) are isostructural with $CsNa_3TiO_4$ ($P\bar{1}$)[30].

SrLiAl$_3$N$_4$ ($I4_1/a$) and SrLiAl$_3$N$_4$ ($P\bar{1}$) are isostructural with NaLi$_3$SiO$_4$ ($I4_1/a$)[31] and KLi$_3$PbO$_4$ ($P\bar{1}$)[32], respectively. SrMg$_3$SiN$_4$ ($P\bar{1}$), SrLiAl$_3$N$_4$ ($I4_1/a$), and CaLiAl$_3$N$_4$ ($P\bar{1}$) are particularly promising because they are predicted to have good phase stability (low $E_{hull}$), a highly rigid crystal structure ($\Theta_D > 600$ K), a HSE band gap within the screening range, and a large splitting in the top two Eu$^{2+}$ 4f bands ($\Delta E_S > 0.1$ eV). BaLiAl$_3$N$_4$ ($P\bar{1}$) and SrLiAl$_3$N$_4$ ($P\bar{1}$) have somewhat higher $E_{hull}$, indicating that they may be more challenging to synthesize. Though CaLiAl$_3$N$_4$ and SrMg$_3$SiN$_4$ (all $P\bar{1}$) have multiple distinct Eu$^{2+}$ activator sites, there is a clear energetic preference (> 15 meV) for one of the sites, and all sites satisfy the criteria for narrow-band emission ($\Delta E_S > 0.1$ eV). We have tested the sensitivity of the screening to changes in the $\Delta E_S$ threshold. As shown in Supplementary Figure S4, we find that there is a distinct separation in the $\Delta E_S$ between broad-band ($\Delta E_S < 0.085$ eV) and narrow-band emitters ($\Delta E_S > 0.1$ eV), and therefore, small variations in the threshold do not affect the classification.

We find that narrow-band emission is indicated only when the Eu$^{2+}$ activator is in one of two highly unusual local environments: (i) an eight-coordinated cuboid-like environment, which is observed in most of the known and predicted narrow-band emitters, including the recently reported SrLiAl$_3$N$_4$:Eu$^{2+}$,[5] and (ii) a nine-coordinated environment, which has thus far been observed only in the narrow-band green emitter β-SiAlON:Eu$^{2+}$. These observations suggest that these highly unusual Eu$^{2+}$ environments are responsible for a crystal field splitting that results in a large gap between the two highest Eu$^{2+}$ 4f bands. To obtain clear evidence of this effect, we extracted the partial charge densities from the computed wavefunctions to determine the orientation of the charge distribution of the individual 4f bands relative to the N$^{3-}$ ligands.

Figure 3 shows the partial charge density of the highest 4$f$ band (at the Fermi level) in the $Eu^{2+}$-activated $CaLiAl_3N_4$ ($I4_1/a$) and β-SiAlON phosphors. We have selected $CaLiAl_3N_4$:$Eu^{2+}$ as a representative phosphor with the $EuN_8$ cuboid-like environment, but have confirmed that the same features are observed in all phosphors with $Eu^{2+}$ in a cuboid-like local environment. We find that the partial charge density for the highest 4$f$ band of the $EuN_8$ environment exhibits a cuboid-like distribution, similar to the atomic $4f_{xyz}$ or $4f_{zx^2-zy^2}$ orbitals, directed along the Eu-N bonds. For the $EuN_9$ environment in β-SiAlON, the partial charge density exhibits a highly symmetric hexagonal distribution, similar to the atomic $4f_{x^3-3xy^2}$, $4f_{y^3-3yx^2}$, $4f_{5yz^2-yr^2}$ or $4f_{5xz^2-xr^2}$ orbitals, with three of the "lobes" aligned with an Eu-N bond each, and the other three "lobes" approximately bisecting a pair of Eu-N bonds each. In contrast, no such alignment with Eu-N bonds is observed in the partial charge density of the lower 4$f$ bands (Supplementary Figures S5 and S6). From a crystal field perspective, a 4$f$ band in either of these special alignments is penalized due to its proximity to the $N^{3-}$ ligands, resulting in a significantly higher energy compared to the other 4$f$ bands.

The implications of our findings go beyond the identification of the five new narrow-band red-emitting phosphors. Previous works have generally attributed narrow-band emission to restricted structural relaxation of the activator in its excited state due to the high symmetry of the cuboid environment.[5] Though we do not rule such constrained relaxation out as a contributing factor, our work suggests that the primary determinant of narrow-band emission is the effect of a highly symmetric crystal field on a highly localized, atomic-like $Eu^{2+}$ 4$f$ orbital, resulting in a large splitting in the top two bands, an effect that can be observed in the *ground-state* band structure. Thus far, only the highly unusual cuboid-like environment and nine-coordinated β-SiAlON environments are observed to have this effect. An open question is whether other coordination environments can be "designed" to induce a similar electronic structure feature for narrow-band emission in $Eu^{2+}$ or other activators, and will be the subject of further investigations. All new narrow-band red-emitting phosphor hosts identified in the high-throughput screening are predicted compounds from the data-mined ionic substitution algorithm. No existing nitrides in the Materials Project database were identified as promising hosts for narrow-band red emission, given that these materials do not have either of the rare environments required for narrow-band emission.

Though the focus of our screening efforts in this work is on narrow-band red-emitting phosphor hosts, the high-throughput first principles screening approach outlined can be readily extended to other emission wavelengths with a desired emission bandwidth. For instance, narrow-band green-emitting phosphors are also required for ultra-efficient solid-state lighting and LED-backlit LCD displays,[12,33] and we have demonstrated that the narrow-band descriptor is applicable to green-emitting phosphors as well (Figure 1c). Conversely, broad-band emission is desired in certain applications to improve CRI, and by suitable inversion of the descriptor and tuning of the screening criteria, broad-band emitting phosphors at a desired wavelength may be identified. Finally, we should note that there are certain limitations inherent in a high-throughput first-principles screening effort. For example, the $Eu^{2+}$ activator concentration is known to affect the emission peak position,[15,25] an effect that is not captured in our screening. Ultimately, the proposed new hosts would need to be synthesized, verified and further optimized experimentally. The fact that several of the proposed hosts are similar in chemistry to known phosphors gives us reasonable confidence that they are synthesizable.

In summary, we have demonstrated that an electronic structure characteristic in narrow-band $Eu^{2+}$-activated phosphors is a large splitting of > 0.1 eV between the two highest $Eu^{2+}$ $4f$ bands. By screening 2,259 ternary, quaternary and predicted nitride compounds on this descriptor and other calculated properties, we have identified five highly promising candidate hosts for $Eu^{2+}$-activated red-emitting phosphors that are predicted to exhibit chemical stability, good thermal quenching resistance and quantum efficiency, and narrow emission bandwidth. We have also shown evidence that narrow-band emission is the result of the crystal-field splitting of the $Eu^{2+}$ $4f$ orbitals in a cuboid or highly symmetrical $EuN_9$ environment, which provides new insights into the emission characteristics of rare-earth dopants in phosphor hosts. The screening strategy in this work provides a general pathway to the discovery of new phosphors with a desired emission colour and bandwidth for solid-state lighting.

## Acknowledgements

This work was supported by the National Science Foundation under Grant No. 1411192. Computational resources from the Extreme Science and Engineering Discovery Environment (XSEDE) are gratefully acknowledged.

## Author contributions

S. P. O. and Z.W. proposed the concept. Z.W. carried out the calculations and analyses, and prepared the initial draft of the manuscript. I.-H. C. performed the GW calculations and charge density analyses. All authors contributed to the discussions and revisions of the manuscript.

## Additional information

Supplementary information is available in the online version of the paper. Reprints and permissions information is available online at www.nature.com/reprints. Correspondence and requests for materials should be addressed to S.P.O.

## Competing financial interests

Zhenbin Wang, Iek-Heng Chu and Shyue Ping Ong have filed a provisional patent on the high-throughput screening methodology and the new phosphors identified in this work.

# References

1.  Uheda, K. *et al.* Luminescence properties of a red phosphor, $CaAlSiN_3:Eu^{2+}$, for white light-emitting diodes. *Electrochem. Solid-State Lett.* **9,** H22 (2006).

2.  Piao, X. *et al.* Preparation of $CaAlSiN_3:Eu^{2+}$ phosphors by the self-propagating high-temperature synthesis and their luminescent properties. *Chem. Mater.* **19,** 4592–4599 (2007).

3.  Li, Y. Q., Hirosaki, N., Xie, R. J., Takeda, T. & Mitomo, M. Yellow-orange-emitting $CaAlSiN_3:Ce^{3+}$ phosphor: structure, photoluminescence, and application in white LEDs. *Chem. Mater.* **20,** 6704–6714 (2008).

4.  Zeuner, M., Hintze, F. & Schnick, W. Low temperature precursor route for highly efficient spherically shaped LED-Phosphors $M_2Si_5N_8:Eu^{2+}$ (M = Eu, Sr, Ba). *Chem. Mater.* **21,** 336–342 (2009).

5.  Pust, P. *et al.* Narrow-band red-emitting $Sr[LiAl_3N_4]:Eu^{2+}$ as a next-generation LED-phosphor material. *Nat. Mater.* **13,** 891–896 (2014).

6.  Pust, P. *et al.* $Ca[LiAl_3N_4]:Eu^{2+}$ —a narrow-band red-emitting nitridolithoaluminate. *Chem. Mater.* **26,** 3544–3549 (2014).

7.  Schmiechen, S. *et al.* Toward new phosphors for application in illumination-grade white pc-LEDs: the nitridomagnesosilicates $Ca[Mg_3SiN_4]:Ce^{3+}$, $Sr[Mg_3SiN_4]:Eu^{2+}$, and $Eu[Mg_3SiN_4]$. *Chem. Mater.* **26,** 2712–2719 (2014).

8.  Xie, R.-J., Hirosaki, N., Li, Y. & Takeda, T. Rare-earth activated nitride phosphors: synthesis, luminescence and applications. *Materials (Basel).* **3,** 3777–3793 (2010).

9. Xie, R.-J., Hirosaki, N., Takeda, T. & Suehiro, T. On the performance enhancement of nitride phosphors as spectral conversion materials in solid state lighting. *ECS J. Solid State Sci. Technol.* **2,** R3031–R3040 (2012).

10. Xie, R.-J. & Bert Hintzen, H. T. Optical properties of (oxy)nitride materials: a review. *J. Am. Ceram. Soc.* **96,** 665–687 (2013).

11. Bardsley, N. *et al. Solid-State Lighting Research and Development Multi-Year Program Plan*. (2014).

12. Phillips, J. M. *et al.* Research challenges to ultra-efficient inorganic solid-state lighting. *Laser Photonics Rev.* **1,** 307–333 (2007).

13. Strobel, P. *et al.* Narrow-band green emitting nitridolithoalumosilicate Ba[Li$_2$(Al$_2$Si$_2$)N$_6$]:Eu$^{2+}$ with framework topology whj for LED/LCD-backlighting applications. *Chem. Mater.* **27,** 6109–6115 (2015).

14. Hirosaki, N. *et al.* Characterization and properties of green-emitting β-SiAlON:Eu$^{2+}$ powder phosphors for white light-emitting diodes. *Appl. Phys. Lett.* **86,** 211905 (2005).

15. Li, Y. Q. *et al.* Luminescence properties of red-emitting M$_2$Si$_5$N$_8$:Eu$^{2+}$ (M=Ca, Sr, Ba) LED conversion phosphors. *J. Alloys Compd.* **417,** 273–279 (2006).

16. Hecht, C. *et al.* SrAlSi$_4$N$_7$:Eu$^{2+}$ − a nitridoalumosilicate phosphor for warm white light (pc)LEDs with edge-sharing tetrahedra. *Chem. Mater.* **21,** 1595–1601 (2009).

17. Dorenbos, P. Thermal quenching of Eu$^{2+}$ 5d–4f luminescence in inorganic compounds. *J. Phys. Condens. Matter* **17,** 8103–8111 (2005).

18. Brgoch, J., DenBaars, S. P. & Seshadri, R. Proxies from ab initio calculations for screening efficient Ce$^{3+}$ phosphor hosts. *J. Phys. Chem. C* **117,** 17955–17959 (2013).

19. Perdew, J. P., Burke, K. & Ernzerhof, M. Generalized gradient approximation made simple. *Phys. Rev. Lett.* **78,** 1396–1396 (1997).

20. Cohen, A. J., Mori-Sanchez, P. & Yang, W. Insights into current limitations of density functional theory. *Science.* **321,** 792–794 (2008).

21. Heyd, J., Scuseria, G. E. & Ernzerhof, M. Hybrid functionals based on a screened Coulomb potential. *J. Chem. Phys.* **118,** 8207 (2003).

22. Heyd, J., Scuseria, G. E. & Ernzerhof, M. Erratum: 'Hybrid functionals based on a screened Coulomb potential' [J. Chem. Phys. 118, 8207 (2003)]. *J. Chem. Phys.* **124,** 219906 (2006).

23. Setyawan, W., Gaume, R. M., Lam, S., Feigelson, R. S. & Curtarolo, S. High-throughput combinatorial database of electronic band structures for inorganic scintillator materials. *ACS Comb. Sci.* **13,** 382–390 (2011).

24. Gál, Z. A., Mallinson, P. M., Orchard, H. J. & Clarke, S. J. Synthesis and structure of alkaline earth silicon nitrides: $BaSiN_2$, $SrSiN_2$, and $CaSiN_2$. *Inorg. Chem.* **43,** 3998–4006 (2004).

25. Duan, C. J. *et al.* Preparation, electronic structure, and photoluminescence properties of $Eu^{2+}$- and $Ce^{3+}/Li^+$-activated alkaline earth silicon nitride $MSiN_2$ (M = Sr, Ba). *Chem. Mater.* **20,** 1597–1605 (2008).

26. Hautier, G. *et al.* Phosphates as lithium-ion battery cathodes: an evaluation based on High-throughput ab initio calculations. *Chem. Mater.* **23,** 3495–3508 (2011).

27. Jain, A. *et al.* Commentary: The Materials Project: A materials genome approach to accelerating materials innovation. *APL Mater.* **1,** 011002 (2013).

28. Hautier, G., Fischer, C., Ehrlacher, V., Jain, A. & Ceder, G. Data mined ionic substitutions for the discovery of new compounds. *Inorg. Chem.* **50,** 656–63 (2011).

29. Bergerhoff, G., Hundt, R., Sievers, R. & Brown, I. D. The inorganic crystal structure data base. *J. Chem. Inf. Model.* **23,** 66–69 (1983).

30. Weiß, C. & Hoppe, R. The new orthotitanate CsNa$_3$[TiO$_4$][1]. *Z. Anorg. Allg. Chem.* **622,** 1715–1720 (1996).

31. Nowitzki, B. & Hoppe, R. Neues uber oxide vom typ A[(TO)$_n$] : NaLi$_3$SiO$_4$, NaLi$_3$GeO$_4$ und NaLi$_3$TiO$_4$. *Rev. Chim. Min.* **23,** 217–230 (1986).

32. Brazel, V. B. & Hoppe, R. 'Fragmentierung' und 'Aggregation' bei Bleioxiden. Ueber das Oligooxoplumbat(IV) K$_2$Li$_6$(Pb$_2$O$_8$). *Z. Anorg. Allg. Chem.* **497,** 176–184 (1983).

33. Wang, L. *et al.* Highly efficient narrow-band green and red phosphors enabling wider color-gamut LED backlight for more brilliant displays. *Opt. Express* **23,** 28707–17 (2015).

34. Tolhurst, T. M. *et al.* Investigations of the electronic structure and bandgap of the Next-generation LED-phosphor Sr[LiAl$_3$N$_4$]:Eu$^{2+}$-experiment and calculations. *Adv. Opt. Mater.* **3,** 546–550 (2015).

35. Fang, C. M., Hintzen, H. T., With, G. de & Groot, R. A. de. Electronic structure of the alkaline-earth silicon nitrides M$_2$Si$_5$N$_8$ (M = Ca and Sr) obtained from first-principles calculations and optical reflectance spectra. *J. Phys. Condens. Matter* **13,** 67–76 (2001).

36. Kresse, G. & Furthmüller, J. Efficient iterative schemes for ab initio total-energy calculations using a plane-wave basis set. *Phys. Rev. B* **54,** 11169–11186 (1996).

37. Blöchl, P. E. Projector augmented-wave method. *Phys. Rev. B* **50,** 17953–17979 (1994).

38. Ong, S. P. *et al.* Python Materials Genomics (pymatgen): A robust, open-source python library for materials analysis. *Comput. Mater. Sci.* **68,** 314–319 (2013).

39. Ong, S. P., Wang, L., Kang, B. & Ceder, G. Li−Fe−P−$O_2$ phase diagram from first principles calculations. *Chem. Mater.* **20,** 1798–1807 (2008).

40. Ong, S. P. *et al.* The Materials Application Programming Interface (API): A simple, flexible and efficient API for materials data based on REpresentational State Transfer (REST) principles. *Comput. Mater. Sci.* **97,** 209–215 (2015).

41. Dudarev, S. L., Savrasov, S. Y., Humphreys, C. J. & Sutton, A. P. Electron-energy-loss spectra and the structural stability of nickel oxide: an LSDA+U study. *Phys. Rev. B* **57,** 1505–1509 (1998).

42. Chaudhry, A. *et al.* First-principles study of luminescence in $Eu^{2+}$-doped inorganic scintillators. *Phys. Rev. B* **89,** 155105 (2014).

43. Hill, R. The elastic behaviour of a crystalline aggregate. *Proc. Phys. Soc. Sect. A* **65,** 349–354 (1952).

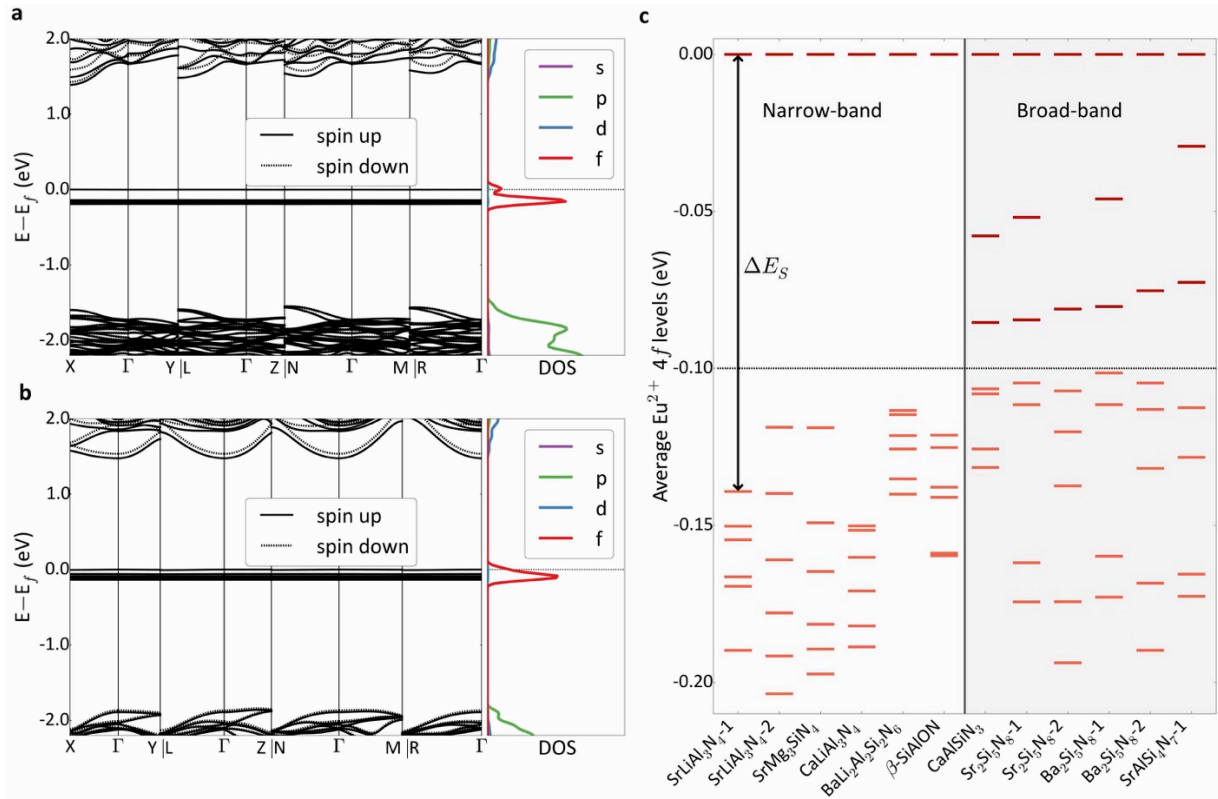

**Figure 1 | Electronic structures for selected (oxy)nitride phosphors**. **a**, Band structure and density of states (DOS, in arbitrary unit) of a narrow-band red-emitting phosphor, $SrLiAl_3N_4:Eu^{2+}$. **b**, Band structure and DOS (in arbitrary unit) of a broad-band red-emitting phosphor, $CaAlSi_3N_4:Eu^{2+}$. **c**, Average $Eu^{2+}$ 4$f$ band levels for five narrow-band and four broad-band phosphors. A numbered suffix (e.g., "-1") is added where necessary to distinguish between distinct $Eu^{2+}$ activator sites in the same host structure, with increasing numbers indicating increasing site energy. The highest 4$f$ band, which lies on the Fermi level, is set at 0 eV for ease of comparison.

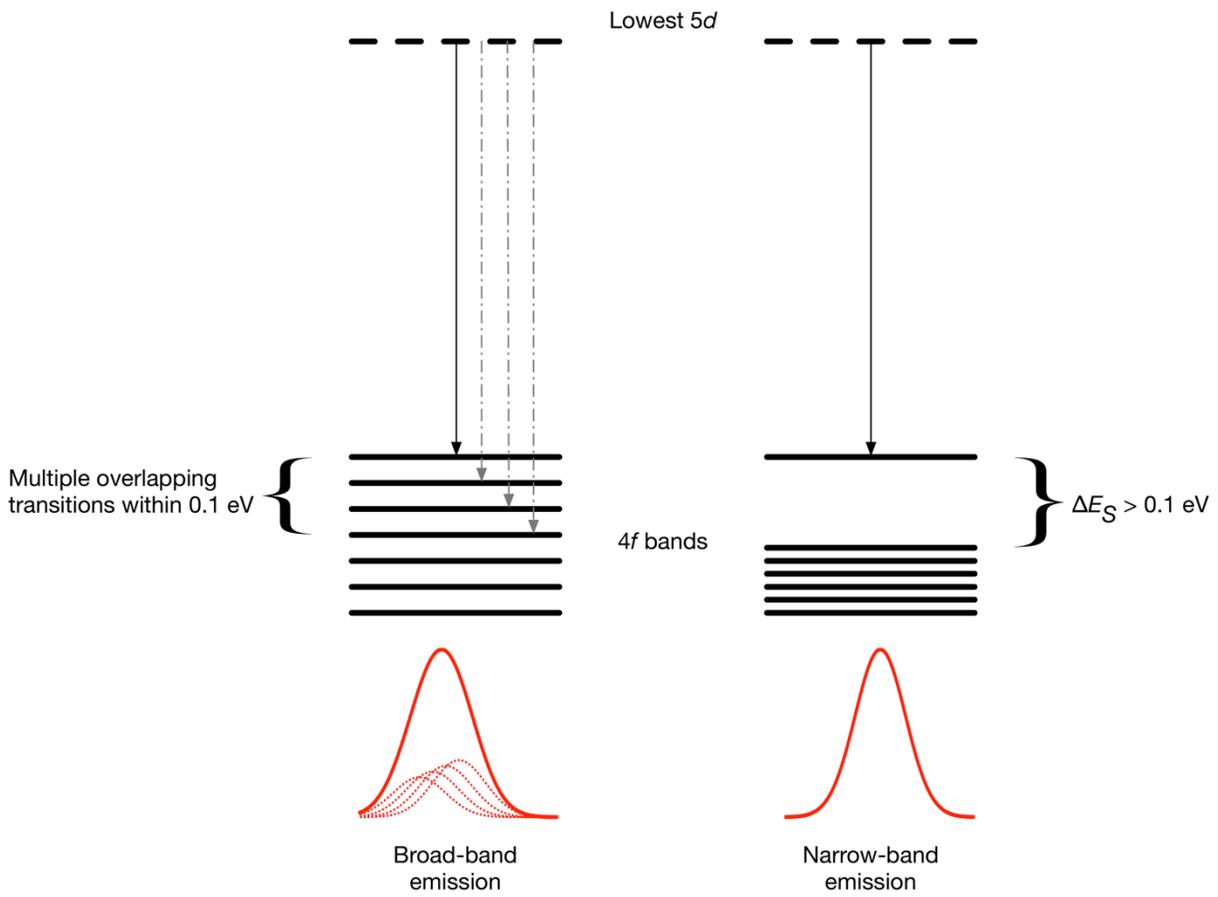

**Figure 2 | Relationship between emission bandwidth and Eu$^{2+}$ 4f band levels.** When there are multiple Eu$^{2+}$ 4f levels within 0.1 eV from the highest band (left), overlapping emissions result in a broad bandwidth. Conversely, a large energy splitting $\Delta E_S$ of > 0.1 eV between the two highest 4f bands (right) result in narrow-band emission.

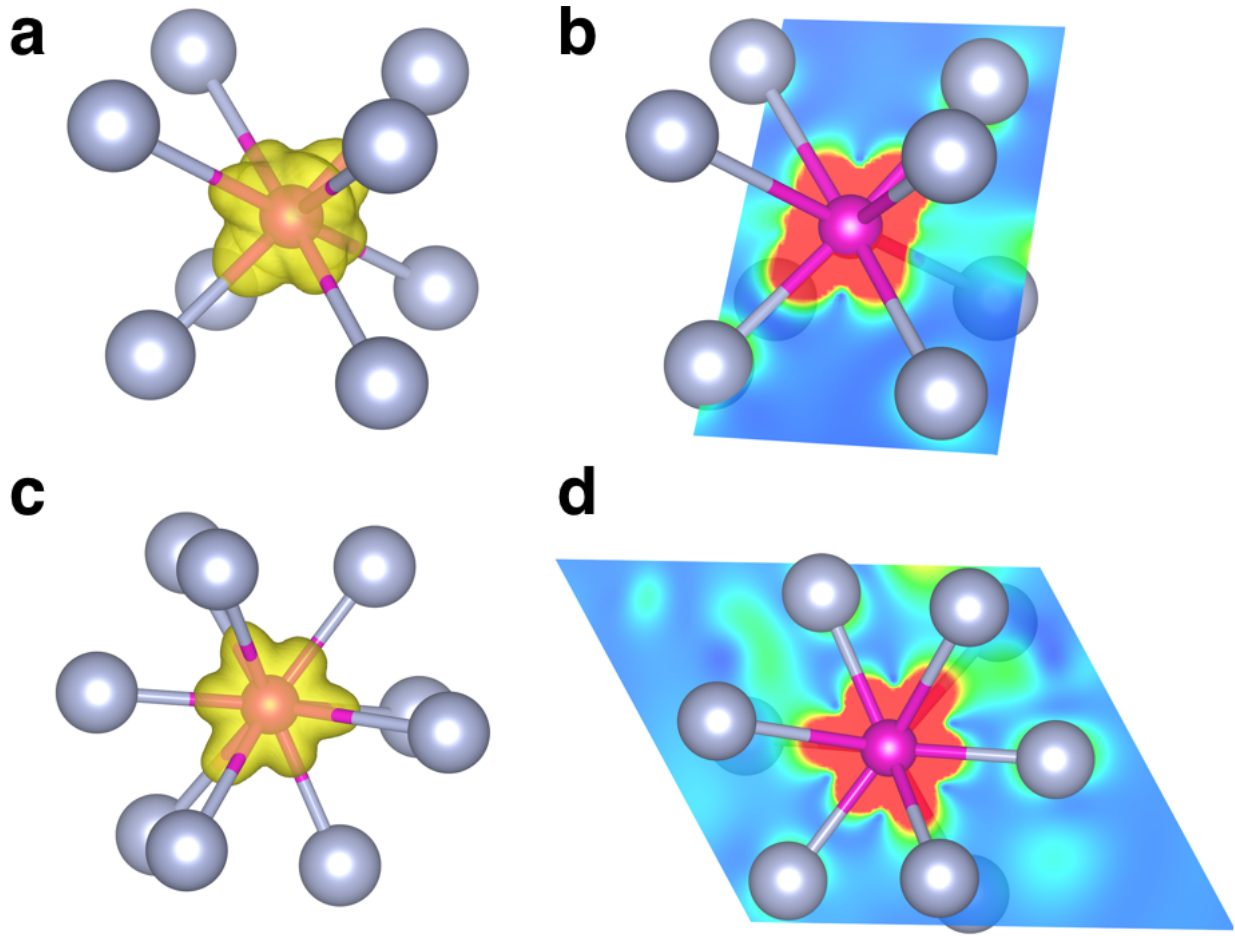

**Figure 3 | Partial charge density of the highest energy Eu$^{2+}$ 4f band. a and c**, Iso-surfaces at 0.0005 $e/a_0^3$ ($a_0$ is Bohr radius) of the partial charge density of the highest energy Eu$^{2+}$ 4f band (at Fermi level) in CaLiAl$_3$N$_4$:Eu$^{2+}$ (cuboid-like EuN$_8$ environment) and in β-SiAlON:Eu$^{2+}$ (highly symmetrical EuN$_9$ environment), respectively. Maroon spheres: Eu, grey spheres: N. **b**. Cross-section of the charge density passing through Eu atom and approximately in the (110) diagonal plane of the cube in CaLiAl$_3$N$_4$:Eu$^{2+}$. **d**. Cross-section of the charge density passing through Eu atom and approximately in the (001) of the EuN$_9$ in β-SiAlON:Eu$^{2+}$.

**Table 1 | Calculated properties of ten known hosts for red-emitting phosphors.** The band gap $E_g$ was calculated using the Perdew-Burke-Ernzerhof (PBE) and screened hybrid Heyd-Scuseria-Ernzerhof (HSE) functionals. Experimental $E_g$, where available, are presented for comparison. $E_{hull}$ and the Debye temperature $\Theta_D$ were calculated using the PBE functional.

| Materials | Space group | $E_{hull}$ (meV) | $E_g$ (eV) PBE | $E_g$ (eV) HSE | $E_g$ (eV) Exp. | $\Theta_D$ (K) |
|---|---|---|---|---|---|---|
| SrLiAl$_3$N$_4$ | $P\bar{1}$ | 0 | 2.97 | 4.47 | 4.56[34], 4.70[5] | 716 |
| SrMg$_3$SiN$_4$ | $I4_1/a$ | 0 | 2.42 | 3.68 | 3.90[7] | 648 |
| CaLiAl$_3$N$_4$ | $I4_1/a$ | 13 | 3.03 | 4.45 | - | 743 |
| CaAlSiN$_3$[a] | $Cc$ | 0 | 3.40 | 4.76 | 4.80[3], 5.0-5.2[2] | 787 |
| Ca$_2$Si$_5$N$_8$ | $Cc$ | 0 | 3.34 | 4.62 | 4.96[15], 4.9[35] | 788 |
| Sr$_2$Si$_5$N$_8$ | $Pmn2_1$ | 0 | 3.20 | 4.40 | 4.67[15], 4.5[35] | 709 |
| Ba$_2$Si$_5$N$_8$[b] | $Pmn2_1$ | 0 | 2.88 | 4.06 | 4.59[15], 4.1[35] | 661 |
| SrAlSi$_4$N$_7$ | $Pna2_1$ | 31 | 3.58 | 4.72 | - | 745 |
| SrSiN$_2$ | $P2_1/c$ | 0 | 2.95 | 4.18 | 4.20[25] | 375 |
| BaSiN$_2$ | $Cmca$ | 0 | 2.92 | 4.03 | 4.10[25] | 360 |

[a] CaAlSiN$_3$ has a disordered structure ($Cmc2_1$). Here, the lowest energy ordered structure ($Cc$) is presented.
[b] Due to the wide variation in the experimentally reported $E_g$ for Ba$_2$Si$_5$N$_8$, $G_0W_0$ calculations were performed for Ba$_2$Si$_5$N$_8$ and BaSiN$_2$. The calculated $G_0W_0$ band gaps for Ba$_2$Si$_5$N$_8$ and BaSiN$_2$ are 4.14 eV and 4.12 eV, respectively, in excellent agreement with the HSE results.

**Table 2 | Calculated properties of eight predicted hosts and the corresponding Eu$^{2+}$-activated phosphors.** The energy above hull $E_{hull}$, band gap $E_g$, Debye temperature $\Theta_D$, relative site energy $E_{site}$ and narrow-band descriptor $\Delta E_S$ were calculated using the PBE functional with a Hubbard $U$ of 2.5 eV for Eu. In addition, more accurate $E_g$ were calculated using the HSE functional. Materials are sorted by stability ($E_{hull}$).

| Materials | Space group | $E_{hull}$ (meV) | $E_g$ (eV) PBE | $E_g$ (eV) HSE | $\Theta_D$ (K) | $E_{site}$ (meV / Eu) | $\Delta E_S$ (eV) |
|---|---|---|---|---|---|---|---|
| **New phosphor hosts** | | | | | | | |
| SrMg$_3$SiN$_4$ | $P\bar{1}$ | 2 | 2.49 | 3.66 | 634 | 0 | 0.124 |
| | | | | | | 17 | 0.115 |
| SrLiAl$_3$N$_4$ | $I4_1/a$ | 3 | 2.93 | 4.12 | 716 | 0 | 0.139 |
| CaLiAl$_3$N$_4$ | $P\bar{1}$ | 14 | 3.00 | 4.28 | 742 | 0 | 0.132 |
| | | | | | | 19 | 0.143 |
| BaLiAl$_3$N$_4$ | $P\bar{1}$ | 28 | 2.46 | 3.64 | 655 | 0 | 0.125 |
| | | | | | | 61 | 0.101 |
| SrLiAl$_3$N$_4$[a] | $P\bar{1}$ | 48 | 2.57 | 3.75 | 704 | 0 | 0.117 |
| **Known phosphor hosts** | | | | | | | |
| SrLiAl$_3$N$_4$[b] | $P\bar{1}$ | 0 | 2.97 | 4.47 | 716 | 0 | 0.139 |
| | | | | | | 17 | 0.119 |
| SrMg$_3$SiN$_4$ | $I4_1/a$ | 0 | 2.42 | 3.68 | 648 | 0 | 0.119 |
| CaLiAl$_3$N$_4$ | $I4_1/a$ | 13 | 3.04 | 4.45 | 743 | 0 | 0.150 |

[a],[b] Though these two materials have the same formula and space group, they are distinct crystal structures. See SI.

## Methods

**Structure relaxation and energy calculations.** Spin-polarized density functional theory (DFT) calculations were performed using the Vienna *ab initio* simulation package (VASP) within the frozen-core projector-augmented wave (PAW) method.[36,37] The generalized gradient approximation (GGA) Perdew-Burke-Ernzerhof (PBE) functional[19] was used for all structural relaxations and energy calculations. A plane wave energy cutoff of 520 eV was used, and the electronic energy and atomic forces were converged to within $10^{-5}$ eV and 0.01 eV/Å respectively. The Brillouin-zone was integrated with a *k*-point density of 1000 per reciprocal atom. All crystal structure manipulations and data analysis were carried out using the Python Materials Genomics package.[38]

**Energy above hull, $E_{hull}$.** The phase stability of predicted materials are estimated by calculating the energy above the linear combination of stable phases in the first principles phase diagram,[39] also known as the energy above hull $E_{hull}$. For phase diagram construction, the energies of all compounds other than those of direct interest in this work were obtained from the Materials Project[27] using the Materials Application Programming Interface.[27,40]

**Host band gap $E_g$, and band structure of $Eu^{2+}$-activated phosphors.** The electronic structures of all host materials were calculated using the PBE functional. For second-tier screening, the screened hybrid Heyd-Scuseria-Ernzerhof (HSE)[21,22] functional was used to obtain more accurate band gaps. To obtain the band structure and 4*f* levels of $Eu^{2+}$-activated phosphors, supercell models were constructed with relatively low $Eu^{2+}$ doping concentrations (< 10%) to mimic experimental doping levels. PBE calculations with a Hubbard $U$[41] parameter of 2.5 eV[42] for Eu was used for these relatively large systems. A Gaussian smearing of 0.05 eV was used for all band structure and density of states calculations. All electronic structure calculations of the $Eu^{2+}$- activated phosphors were performed without spin-orbital coupling (SOC), as the difficult convergence of SOC calculations makes them unsuitable for a high-throughput screening effort. Nevertheless, we performed SOC analyses on the five new narrow-band red-emitting phosphors identified, and confirmed that the relevant electronic structure feature, *i.e.*, a large splitting in the top two 4*f* bands, does not change significantly with the inclusion of SOC.

**Debye temperature $\Theta_D$.** The Debye temperature ($\Theta_D$) was calculated using the quasi-harmonic model.[18] The elastic tensor was calculated with more stringent electronic convergence criterion of $10^{-6}$ eV per formula unit, and the elastic moduli were calculated using the Voigt-Reuss-Hill (VRH) approximation.[43]

**High throughput screening.** Supplementary Figure S7 provides a summary of the high-throughput screening approach employed in this work, which is tiered based on the relative computational expense to calculate the screening properties. An initial list of candidate structures was first generated from all existing ternary and quaternary nitrides in the Materials Project database,[27] which contains the pre-computed data for all ordered known inorganic crystals from the Inorganic Crystal Structure Database (ICSD). To augment this dataset, a prediction of novel nitridosilicate and nitriodoaluminate structures with formula $A_xB_yC_zN_n$ (A = Ca/Sr/Ba, B = Li/Mg, C = Al/Si) was carried out by applying the data-mined ionic substitution algorithm proposed by Hautier et al.[28] on all crystal structures in the ICSD.[29] In the first screening stage, all materials with $E_{hull}$ > 50 meV, indicating that they are unlikely to be stable, were screened out. This was followed by further screening for candidates with 2.42 eV < PBE $E_g$ < 3.58 eV. Finally, the 4*f* band levels of the $Eu^{2+}$-activated host were calculated to find narrow-band emitters with $\Delta E_S$ > 0.1 eV. The HSE $E_g$ and Debye temperature $\Theta_D$ were then computed for all materials that remain.